\documentclass[aps,prl,reprint,groupedaddress]{revtex4-1}

\usepackage{graphicx}
\usepackage{amsmath}
\usepackage{amssymb}

\begin{document}

\preprint{}

\title{Polarization-dependent property of dipole-dipole
interaction mediated by localized surface plasmon of an Ag nanosphere}
\author{Yun-Jin Zhao$^{1}$,Yong-Gang Huang$^{1,2,\ddag}$, Hong Yang$^{1}$, Xiao-Yun Wang$^{1}$, Ke Deng$^{1}$, Zheng Zhang$^{1}$,He-Ping Zhao$^{1}$, and Zhengyou Liu$^{2,3,}$}
\email[$^\ddag$]{huang122012@163.com}
\email[$^*$]{zyliu@whu.edu.cn}
\address{$^1$College of Physical Science and Mechanical Engineering, Jishou University,
Jishou 416000, China }
\address{$^2$Key Laboratory of Artificial Micro- and Nano-structures of Ministry of
Education and School of Physics and Technology, Wuhan University, Wuhan
430072, China}
\address{$^3$Institute for Advanced Studies, Wuhan University, Wuhan 430072, China}
\date{\today}

\begin{abstract}
The effect of the dipole polarization on the quantum dipole-dipole interaction near an Ag nanosphere (ANS) is investigated. A theoretical formalism in terms of classical Green function is developed for the transfer rate and the potential energy of the dipole-dipole interaction (DDI) between two polarized dipoles. It is found that a linear transition dipole can couple to a left circularly polarized transition dipole much stronger than to a right circularly polarized transition dipole. This polarization selectivity exists over a wide frequency range and is robust against the variation of the dipoles¡¯ position or the radius of the ANS. In contrast, a right circularly polarized transition dipole, can change sharply from coupling strongly to another right circularly polarized dipole to coupling strongly to a left circularly polarized dipole with varying frequency. However, if the two dipoles are placed in the same radial direction of the sphere, the right circularly polarized transition dipole can only couple to the dipole with the same polarization while not to the left circularly polarized transition dipole. These findings may be used in solid-state quantum-information processing based on the DDI.
\end{abstract}

\pacs{42.50.Ct, 34.20.-b, 37.30.+i}

\maketitle
\emph{Introduction. }Dipole-Dipole interaction (DDI) between ¡®atoms¡¯ is one of the most fundamental two-body interactions in quantum electrodynamics\cite{Principles_of_nano-optics,
Coherence_in_Spontaneous_Radiation_ProcessesPhysRev.93.99,
Radiation_from_an_N_Atom_SystemPhysRevA2_889,
PhysRevLett.100.243201,
PhysRevLett.93.153001,
Cooperative_Phenomena_in_Resonant_Electromagnetic_PropagationPhysRevA.2.1730,
Retardation-in-the-resonant-interaction-of-two-identical-atomsPhysRevA.10.1096,
Quantum_electrodynamics_of_resonant_energy_transferPhysRevB.49.8751,
Resonant-dipole-dipole-interaction-in-the-presence-of-dispersingPhysRevA.66.063810}. It stems from photon exchange where a photon, both real and virtual, emitted by one atom could be absorbed by the other atom. The transfer rate (incoherent part) and the DDI potential energy (coherent part) are determined by the rates of photon emission, transmission and absorption. Much effort has been devoted to controlling or changing the DDI by the electromagnetic environment\cite{PhysRevA.52.4083,
PhysRevA.57.667,
PhysRevA.52.2835,
PhysRevA.56.5135,
PhysRevLett.82.4118,
Entanglement-of-Two-Qubits-MediatedPhysRevLett.106.020501,
Quantum-emitters-coupled-to-surface-plasmons-nanowirePhysRevB.82.075427,
Dipole-dipole-shiftPhysRevB.84.075419,
Entanglement-generation-between-two-atoms-via-surface-modesPhysRevA.84.032334,
Robust-to-loss-entanglement-generation,
A-revisitation-of-the-Forster-energy-transfer-near-metallic-spherical,
Dissipation-driven-entanglement-mediated-nanoantennasPhysRevB.89.235413,
Entanglement-of-two-three-four-plasmonicallyPhysRevB.92.125432,
Nanoshell-mediated-robust-entanglementPhysRevA.93.022320,
Dipole-dipole-interactionPhysRevA.85.053827,
Surface-plasmon-polariton-assisted-dipole-dipole-interaction-near-metal-surfaces,
Long-range-dipole-dipole-interactionPhysRevB.93.245439} (e.g., cavity, waveguide and plasmonic nanostructure). Many novel quantum phenomena have been predicted\cite{Building_one_molecule_from_a_reservoir_of_two_atoms2018,
miroshnychenko2009observation,
Observing-the-Dynamics-Dipole-Mediated-Energy-Transport,
PhysRevLett.82.1060,
PhysRevLett.87.037902,
PhysRevLett.87.037901,
PhysRevLett.74.4083,
PhysRevA.79.012309,
PhysRevLett.100.113003,
PhysRevLett.95.243602,
PhysRevA.53.R35,
PhysRevA.65.043813,
Fluorescence_Energy_Transfer_as_Spectroscopic_Ruler,
Fluorescence_resonance_energy_transfer,
Dissipation-driven-generationPhysRevB.84.235306,
Two-qubit-correlations-via-a-periodic-plasmonic,
Resonance_interaction_energy_between_two_entangled_atoms_in_a_photonic_bandgap_environment2018,
Controlling_resonance_energy_transfer_in_nanostructure_emitters_by_positioning_near_a_mirror2017,
Transfer-of-energy-between-a-pair-of-molecules-near-a-plasmonic-core-shell-nanoparticle,
Quantum-coherence-and-quantum-correlation,
Squeezed-Light-from-Entangled-Nonidentical,
Superradiance-subradiance-PhysRevA.82.023827,
PhysRevA.42.5695,
0295_5075_9_1_015,
1674_1056_24_2_024205}, such as controlled reaction for two atoms to form one molecule, visualize coherent dipole coupling in real space at the single-molecular level, quantum entanglement preparation and quantum information processing, dipole blocked, cooperative radiation, enhanced and inhibited F\"orster energy transfer and so on.

In the previous investigations, the matrix element of the transition dipole moment has been chosen to be real vector. In addition to the strength, its orientation has also been investigated\cite{The_orientational_freedom_of_molecular_probes,
doi:10.1021/bi00616a032}. On the other hand, the dipole can be elliptically polarized (e.g., quantum dot or atom with transition dipole moment element between the Zeeman levels with different magnetic quantum number), where the matrix element is a complex vector (see a recent review \cite{Interfacing-single-photonsRevModPhys.87.347}). Very recently, this has been extensively investigated in the area of photon-emitter coupling, many novel phenomena associated with complex transition dipole moment element have been demonstrated\cite{A-quantum-gateReisererKalb-32,
Observation-of-entanglement-GaoFallahi-33,
Polarization-Engineering-in-Photonic-CrystalPhysRevLett.115.153901,
Directional-spontaneous-emissionPhysRevA.92.043819,
Charged-quantum-dotPhysRevB.93.241409,
Chiral-nanophotonic-waveguide-interface-based-on-spin-orbit-interactionPetersen67,
Chirality-of-nanophotonic-waveguideColesPrice-34,
Macroscopic-rotation-of-photon-polarization,
A-quantum-phase-switch,
Deterministic-photon¨Cemitter-coupling,
Controlling-circular-polarizationPhysRevB.92.205309,
Spontaneous-emission-of-a-two-level-atomPhysRevA.93.043828,
Theory-of-optical-spinPhysRevB.92.115305}. For example, unidirectional emission, macroscopic rotation of photon polarization, photon-photon interaction mediated by spin, trapping atoms by vacuum forces and so on. Thus, it is necessary and important to take the dipole polarization into account.

In this work, we demonstrate the effect of the dipole polarization on the DDI between two atoms locating around an ANS. Within the framework of quantization of the macroscopic electromagnetic field, we develop a general formalism that the transfer rate and the potential energy of the DDI are expressed by the classical photon Green tensor. In contrast with the real matrix element of the transition dipole moment, both the transfer rate and the DDI potential energy are a mixture of the imaginary part and the real part of the photon Green tensor for complex matrix element of transition dipole moment. Since the real part and the imaginary part of the photon Green function exhibit quite different characteristics, this mixing is vital and brings us a new degree of freedom to tailor the DDI, which is forbidden for real transition dipole matrix element. To show this, we keep the polarization of one atom (atom B) invariant and consider two different kinds of polarization for the other atom (i.e., right circularly polarization and left circularly polarization). It is found that there is a huge difference between the two kinds of polarization over a wide frequency range for both of the transfer rate and the DDI potential energy. We also find that this polarization selective DDI is robust against the fine tuning of the position of the atom or the radius of the ANS. Besides, other interesting phenomena such as polarization dependent switch effect and deterministic coupling are demonstrated. In addition to the implications in fundamental studies, this sensitive polarization-dependent characteristic of the DDI should lead to important applications (e.g., quantum information processing).

\emph{Theory.} By macroscopic QED formalism, the multipolar-coupling Hamiltonian in the
electric dipole approximation, for two two-level `atoms' located in the presence of dispersing and absorbing bodies, reads \cite{Resonant-dipole-dipole-interaction-in-the-presence-of-dispersingPhysRevA.66.063810}
\begin{align}
H &=H_{0}+H_{I} \nonumber\\
H_{0}&=\int d\mathbf{r}\int_{0}^{\infty }d\omega \hbar \omega \text{ }\hat{\mathbf{f}}^{\dagger }\left(\mathbf{r},\omega \right) \cdot \hat{\mathbf{f}}
\left( \mathbf{r},\omega \right) +\underset{i=A,B}{\sum}\hbar \omega
_{i}|e_{i}\rangle \langle e_{i}|, \nonumber\\
H_{I} &=-\underset{i=A,B}{\sum }\int_{0}^{\infty }d\omega \lbrack
(\mathbf{d}_{i}\sigma _{i}^{-}+\mathbf{d}_{i}^{\ast }\sigma _{i}^{+})\cdot \hat{\mathbf{E}}\left(
\mathbf{r}_{i},\omega \right) +\mathbf{H.c}.].
\label{eq:Hamiltonian}
\end{align}

Here, $H_{0}$ is the noninteracting Hamiltonian for the field and the two `atoms'. $H_{I}$ is the atom-field
interaction part, which includes the counter-rotating term.  $\omega _{A}$ ($\omega _{B}$) and $\mathbf{r}_{A}$ ($\mathbf{r}_{B}$) are
the transition frequency and position of atom A (B) , respectively. $%
|e_{A}\rangle $ ($|e_{B}\rangle $) and $|g_{A}\rangle $ ($|g_{B}\rangle$) are the excited and ground states of atom A(B).  $\mathbf{d}_{i}=\langle g_{i}|%
\hat{\mathbf{d}}_{i}|e_{i}\rangle $ is the element of the transition dipole moment, which is complex vector in general. $\hat{\mathbf{f}}(\mathbf{r},\omega)$ is the
bosonic vector field annihilation operator for the elementary excitations of
the environment, which obeys $[\hat{f}_i(\mathbf{r},\omega),\hat{f}_j^\dag(\mathbf{r^{'}},\omega^{'})]=
\delta_{ij}\delta(\mathbf{r}-\mathbf{r^{'}})\delta(\omega-\omega^{'})$,
$[\hat{f}_i(\mathbf{r},\omega),\hat{f}_j(\mathbf{r^{'}},\omega^{'})]
=[\hat{f}_i^\dag(\mathbf{r},\omega),\hat{f}_j^\dag(\mathbf{r^{'}},\omega^{'})]=0$.
The electric field operator at the position of the atom is given by
$\hat{\mathbf{E}}(\mathbf{r},\omega )=i\sqrt{\hbar /\pi \varepsilon _{0}}\int
d\mathbf{r}^{\prime }\sqrt{\varepsilon _{I}(\mathbf{r}^{\prime },\omega )}\mathbf{G}(\mathbf{r},\mathbf{r}^{\prime },\omega )\cdot \hat{\mathbf{f}}(\mathbf{r}^{\prime },\omega )$ with $\varepsilon _{I}(\mathbf{r}^{\prime },\omega )$  the imaginary part
of the complex dielectric function $\varepsilon (\mathbf{r},\omega )=\varepsilon _{R}(\mathbf{r}^{\prime },\omega )+i\varepsilon _{I}(\mathbf{r}^{\prime },\omega )$. Further more, the Green tensor $\mathbf{G}(\mathbf{r},\mathbf{r}^{\prime },\omega )$\
satisfies
$\nabla \times \nabla \times \mathbf{G}(\mathbf{r},\mathbf{r}^{\prime };\omega )-\varepsilon (\mathbf{r},\omega )%
\omega ^{2}/c^{2}\mathbf{G}(\mathbf{r},\mathbf{r}^{\prime };\omega )=\omega ^{2}/c^{2}\mathbf{I}\delta (\mathbf{r}-\mathbf{r}^{\prime })
$, which can be solved semi-analytically or numerically (see Ref. \cite{Zhao:18} and references therein).

Similar to Ref. \cite{Retardation-in-the-resonant-interaction-of-two-identical-atomsPhysRevA.10.1096,
A-revisitation-of-the-Forster-energy-transfer-near-metallic-spherical,
Charged-quantum-dotPhysRevB.93.241409,
Chiral-route-to-spontaneous-entanglement-generationPhysRevB.92.155304}, we assume initially that atom A is in the excited state, atom B is in the ground state and the field is in vacuum. In this case, the states of
interest are $|a\rangle =|e_{A},g_{B},0\rangle $, $|b\rangle
=|g_{A},e_{B},0\rangle $, $|\mathbf{C}\left(\mathbf{r},\omega \right)\rangle
=\hat{\mathbf{f}}^{\dagger}\left( \mathbf{r},\omega
\right)|g_{A},g_{B},0 \rangle$ and $|\mathbf{D}(\mathbf{r},\omega)\rangle=\hat{\mathbf{f}}^{\dagger}\left( \mathbf{r},\omega\right)|e_{A},e_{B},0\rangle $, which form an approximate complete set of states \cite{Retardation-in-the-resonant-interaction-of-two-identical-atomsPhysRevA.10.1096}. We will see later that this approximation is equivalent to the perturbative expansion of the level shift operator in powers of $H_I$ to the second order which neglects the two-photon and many-photon process ( see Eq. (\ref{eq:perturbativeRz}) ).

The initial state is denoted as $|a\rangle$ and the state vector of the system evolves as $|\Psi(t)\rangle \equiv U(t)|a\rangle$, where $U(t)$ is the evolution operator and reads
\begin{equation}
U(t)=\frac{1}{2\pi i}\int_{-\infty}^{+\infty }d\omega\left[G^{-}(\omega )-G^{+}(\omega )\right]\exp (-i\omega t).
\label{eq:evolution}
\end{equation}

Here, $G^{\pm }(\omega )=\lim_{\eta \rightarrow 0^{+}}G(z=\omega \pm i\eta )$,
with the resolvent operator defined by $G(z)=(z-H/\hbar )^{-1}$ \cite{Atom-Photon-Interactions-Basic-Processes-and-Application,Quantum-statistical-theories-of-spontaneous-emissionAgarwal1974}, which can be resolved by the projection operator method. Let $\alpha $\ be a subspace subtended by the eigenvector of $H_{0}$ \{$|a\rangle $ $|b\rangle $\}, then $P=|a\rangle
\langle a|+|b\rangle \langle b|$ is the projector on to this subspace and $%
Q=I-P$ is the projector on to the supplementary subspace of $\alpha$. By
using the identity $(z-H/\hbar )G(z)=1$, we find $PG(z)P =\hbar\frac{P}{\hbar z-PH_{0}P-\hbar PR(z)P}$ with the level shift operator defined by $\hbar R(z)=H_{I}+H_{I}\frac{Q}{\hbar z-QH_{0}Q-QH_{I}Q}H_{I}$. The perturbative expansion for the level shift operator in powers of $H_I$ reads
\begin{align}
\hbar R(z) & =H_{I}+H_{I}\frac{Q}{\hbar z-H_{0}}H_{I}+\notag\\
&+H_{I}\frac{Q}{\hbar z-H_{0}}H_{I}\frac{Q}{\hbar z-H_{0}}H_{I}....
\label{eq:perturbativeRz}
\end{align}

Thus, for our system (see Eq. (\ref{eq:Hamiltonian}) with initial state $|e_{A},g_{B},0\rangle $), to restrict the state of the system in the space spanned by \{$|a\rangle , |b\rangle, |\mathbf{C}\left(\mathbf{r},\omega \right)\rangle,|\mathbf{D}\left(\mathbf{r},\omega \right)\rangle $\} is equivalent to take the first two terms in Eq. (\ref{eq:perturbativeRz}). It should be noted that the single photon effect is completely included by this method.

With some calculation, the matrix elements for the resolvent operator can be solved. Explicitly, $G_{aa}(z) =(z-\omega_{B}-R_{bb}(z))/\Xi$ and $G_{ba}(z) =R_{ba}(z)/\Xi$, where
$\Xi=\left[ z-\omega_{A}-R_{aa}(z)\right] \left[ z-\omega_{B}-R_{bb}(z)\right]
-R_{ab}(z)R_{ba}(z)$ with $F_{ij}(z)$ defined by $\langle i|F(z)|j\rangle$.
$R_{ii}(z)$ and $R_{ij}(z)$ are the local coupling between the atom and the field and the dipole-dipole coupling between atoms $i$ and $j$, respectively. The transfer rate $\Gamma
_{ij}(z)$ and the potential energy $\Delta _{ij}(z)$ of the DDI are related to
the imaginary part (incoherent) and the real part (coherent) of the matrix element of the level shift operator $%
\lim_{\eta \rightarrow 0^{+}}R_{ij}(z\pm i\eta)=\Delta _{ij}(z)\mp i\Gamma _{ij}(z)/2$, respectively.

By utilizing the Kramers-Kronig relation of the Green function, one obtains
\begin{equation}
\hbar R_{ba}(z+i\eta)=\frac{1}{\pi\varepsilon_{0}}\mathbf{d}_{B}^{\ast}\cdot
\mathbf{M}_{ba}\cdot\mathbf{d}_{A},
\label{Rba}
\end{equation}
where
\begin{widetext}
\begin{align}
\mathbf{M}_{ba}  &  =\int_{-\infty}^{\infty}d\omega\frac{\operatorname{Im}%
\mathbf{G}(\mathbf{r}_{B},\mathbf{r}_{A},\omega)}{z-\omega+i\eta}+  \int_{0}^{\infty}d\omega\operatorname{Im}\mathbf{G}(\mathbf{r}%
_{B},\mathbf{r}_{A},\omega)(\frac{1}{z+\omega+i\eta}+\frac{1}{z-\omega
_{A}-\omega_{B}-\omega+i\eta})\nonumber \\
&=-\pi\mathbf{G}(\mathbf{r}_{B},\mathbf{r}_{A},z)+\mathbf{R}_{1}^{c}(z)+\mathbf{R}_{2}^{c}(z)+\mathbf{R}_{3}^{c}(z),
\end{align}
\end{widetext}
with the three correcting parts written by
\begin{small}
\begin{subequations}
\begin{align}
\mathbf{R}_{1}^{c}(z) &  =-i\pi\int_{0}^{\infty}d\omega\operatorname{Im}%
\mathbf{G}(\mathbf{r}_{B},\mathbf{r}_{A},\omega)\delta(z+\omega),\\
\mathbf{R}_{2}^{c}(z) &  =-i\pi\int_{0}^{\infty}d\omega\operatorname{Im}%
\mathbf{G}(\mathbf{r}_{B},\mathbf{r}_{A},\omega)\delta(z-\omega_{A}-\omega
_{B}-\omega),\\
\mathbf{R}_{3}^{c}(z)&=-2\wp\int_{0}^{\infty}d\omega
\operatorname{Im}\mathbf{G}(\mathbf{r}_{B},\mathbf{r}_{A},\omega
)[\frac{z-\overline{\Omega}}{(\overline{\Omega}+\omega)^{2}-(z-\overline
{\Omega})^{2}}]. \label{R3}
\end{align}
\end{subequations}
\end{small}

Here, $\overline{\Omega}=0.5(\omega_{A}+\omega_{B})$ is the average transition frequency for the two `atoms' and $\wp$ stands for the principle integration. This is the general results for the matrix element of the level shift operator. If $\omega_{A}=\omega_{B}=\overline{\Omega}$, the above three correction parts vanish at $z=\overline{\Omega}$, $\mathbf{R}_{1}^{c}(\overline{\Omega})=\mathbf{R}_{2}^{c}(\overline{\Omega})=\mathbf{R}_{3}^{c}(\overline{\Omega})=0$ and Equation (\ref{Rba}) is the same as that found in the literature ( for example, Eq. (39) in Ref. \cite{PhysRevA.65.043813}). Except for the case of ultra-strong coupling, where the real parts of the zeros for equation $\Xi=0$ ( the denominator for matrix element of the resolvent operator $G_{ij}$ ) may be negative, the three correction parts ( $\mathbf{R}_{1}^{c}(z)$, $\mathbf{R}_{2}^{c}(z)$ and $\mathbf{R}_{3}^{c}(z)$ ) are small and can be neglected. For our ANS system, which is also investigated in Ref. \cite{Zhao:18,Spontaneous-emission-spectraPhysRevB.85.075303,1674_1056_24_2_024205}, the energy level shift and dipole-dipole shift are much more less than the transition frequency where the dipole strength is $d=24D$ and its distance to the sphere surface is $2nm$. In this case, the real parts of the zeros of equation $\Xi=0$ are around $\overline{\Omega}$ and the matrix elements of the resolvent operator $G_{aa}(z)$ and $G_{ba}(z)$ are peaked around $\overline{\Omega}$. Thus, only the positive $z$ should be considered and we get $\mathbf{R}_{1}^{c}(z) = 0$ and $\mathbf{R}_{2}^{c}(z)= 0$. For $\mathbf{R}_{3}^{c}(z)$, the integrand in the bracket of Eq. (\ref{R3}) is very small for $z$ around $\overline{\Omega}$ and $\operatorname{Im}\mathbf{G}(\mathbf{r}_{B},\mathbf{r}_{A},\omega
)$ can be described by the sum of many Lorentzian functions \cite{
Adiabatic-passage-mediated-by-plasmonsPhysRevB.93.045422,Mode-selective-quantization-and-multimodal-effective-modelsPhysRevA.94.023818}. Thus, $\mathbf{R}_{3}^{c}(z)$ is very small and can be neglected in our nanosphere system, which has been numerically verified by us. Then the transfer rate and the DDI potential energy reads
\begin{subequations}
\label{eq:gamadeltazhongyao}
\begin{align}
\Gamma_{ij}(\omega) &  =\frac{2}{\hbar\varepsilon_{0}}\operatorname{Im} [\mathbf{d}_{i}^{\ast}\cdot \mathbf{G}(\mathbf{r}_{i},\mathbf{r}_{j},\omega)\cdot \mathbf{d}_{j}],\\
\Delta_{ij}(\omega) &  =-\frac{1}{\hbar\varepsilon_{0}%
}\operatorname{Re} [\mathbf{d}_{i}^{\ast}\cdot \mathbf{G}(\mathbf{r}_{i},\mathbf{r}_{j},\omega)\cdot \mathbf{d}_{j}].\label{eq:deltaij}
\end{align}%
\end{subequations}

The above equations are the main results of our theory. Following the similar procedure, we can extend the above results to the case for $i=j$ with the Green function in Eq.(\ref{eq:deltaij}) being replaced by the scatter Green tensor. The divergent part of the homogeneous-medium contribution can be absorbed into the transition frequency $\omega_i$. For $i=j$, $\Gamma_{ii}(\omega_{i})$ and $\Delta_{ii}(\omega_{i})$ are the spontaneous emission rate and Lamb shift for atom $i$ in the weak coupling regime. $\Gamma_{ii}$ ($\Delta_{ii}$) is related to the imaginary (real) part of the Green function and there is no mix between the real part and the imaginary part. For example, different polarizations of the dipole moment mix different components of the imaginary part of the Green function for $\Gamma_{ii}$. The same is true for $\Delta_{ii}$. However, for $i\neq j$ and complex transition dipole moment element, both the transfer rate $\Gamma_{ij}(\omega)$ and the potential energy $\Delta_{ij}(\omega)$ are a mix of the real part and the imaginary part of the Green function.  For example, if the dipole moment of atom $j$ is linear polarized along the \emph{x} axis ($\mathbf{d}_j=\hat{e}_x$) and atom $i$ is right circularly polarized ($\mathbf{d}_i=\hat{e}_x+i\hat{e}_y$), where $\hat{\mathbf{e}}_{x}$, $\hat{\mathbf{e}}_{y}$ and $\hat{\mathbf{e}}_{z}$ are the unit vectors along
the \emph{x}-axis, \emph{y}-axis and \emph{z}-axis respectively,  $\Gamma_{ij}(\omega)=2/(\hbar\varepsilon_{0})[\hat{e}_x\cdot \operatorname{Im} \mathbf{G}(\mathbf{r}_{i},\mathbf{r}_{j},\omega)\cdot \hat{e}_x-\hat{e}_y\cdot \operatorname{Re} \mathbf{G}(\mathbf{r}_{i},\mathbf{r}_{j},\omega)\cdot \hat{e}_x]$ , which is different from the real dipole moment case or the $i=j$ case. Since the real part and the imaginary part of the Green function exhibit very different characteristics in nano-structures, the polarization of the dipole will take great effect in the DDI.

We emphasize that Eq. (\ref{eq:gamadeltazhongyao}) is the general quantum description of the DDI, even though the Green tensor is classical. They can be applied for atoms located in any lossy and inhomogeneous structure. In addition, they provide suitable description of the DDI for any type of electric transition dipole moment, either complex or real.
\begin{figure}[htbp]
\includegraphics[width=6.5cm]{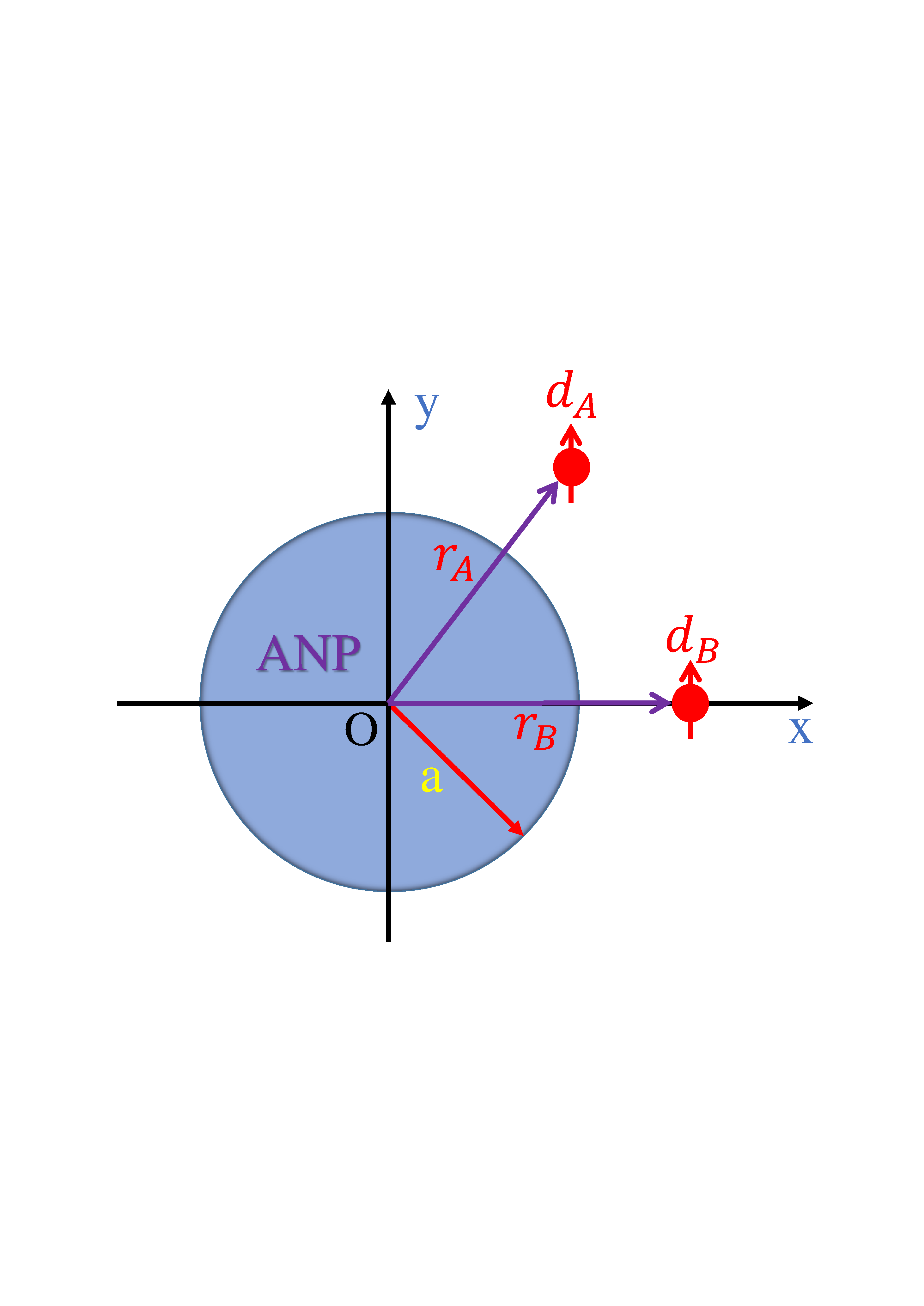}
\caption{(Color online). Schematic diagram of the system. An ANP with radious $a$ is located at the origin in vacuum. Atoms A and B
are around the ANS in the $xy$-plane and can interact with each other mediated by the ANS. Their transition dipole moment are $\mathbf{d}_{A}$\ and $\mathbf{d}_{B}$. The positions of atoms A and B are $\mathbf{r}_{A}$ and $\mathbf{r}_{B}$.}
\label{fig1}
\end{figure}

\emph{Model.} The schematic diagram of the system is illustrated in Fig. 1. There is an ANS with radius $a$ at the coordinate origin. The permittivity of the ANS is given by the Drude model, $\varepsilon_{m}=\varepsilon_{\infty}-\omega_{m}^{2}/(\omega^{2}+i\gamma_{m}\omega)$ with $\varepsilon
_{\infty}=6$, $\omega_{m}=7.90eV$ and $\gamma=51meV$\cite{Spontaneous-emission-spectraPhysRevB.85.075303,
Mode-selective-quantization-and-multimodal-effective-modelsPhysRevA.94.023818}. The two atoms are located on the $xoy$-plane and their positions are denoted by $\mathbf{r}_{A}$\ and $\mathbf{r}_{B}$. Except for special statement, we set $a=20nm$, $\mathbf{r}_{B}=(22nm,0)$ and denote the position of atom A as $\mathbf{r}_{A}=(x_{A},y_{A})$. The transition dipole
moment for atom $i$ is $\mathbf{d}_{i}=u_i\mathbf{\hat{u}}_{i}$, with $\mathbf{\hat{u}}_{i}$ the complex unitary vector. Since we are much interest in the polarization effect, we set $u_A=u_B=u$. The transfer rate and the potential energy of the DDI are normalized by the vacuum radiation rate $\Gamma_{0}(\omega
)=u^{2}\omega^{3}/3\pi\varepsilon_{0}\hbar c^{3}$. For simplicity, we use
$\Gamma_{ij}$\ and $\Delta_{ij}$\ for $\Gamma_{ij}(\omega)$\ and $\Delta
_{ij}(\omega)$, respectively.

\emph{Results.} Owing to the inversion symmetry of the ANS about the $xy$-plane, we find
$G_{zj}=G_{jz}=0$ with $G_{zj}=\hat{\mathbf{e}}_{z}\cdot \mathbf{G}\cdot\hat{\mathbf{e}}_{j}$ ($j=x, y$). This means that a $z$-polarized
dipole will only interact with a dipole containing $z$-component and not
interact with any dipoles polarized in the $xy$-plane, which is simple. Thus, we
first consider the case that both dipoles are polarized in the $xy$-plane.
Figure 2 shows the polarization-dependent characteristics for the potential
energy $\Delta_{AB}$ and the transfer rate $\Gamma_{AB}$ between a linear
polarized transition dipole $\mathbf{\hat{u}}_{B}=\hat{\mathbf{e}}_{x}$\ and a
circularly polarized transition dipole $\mathbf{\hat{u}}_{A}=\mathbf{\hat{u}%
}_{A}^{\pm}=(\hat{\mathbf{e}}_{x}\pm i\hat{\mathbf{e}}_{y})/\sqrt{2}$. In order to describe the
different coupling property for the two polarization state of atom A, we define
the contrast of the transfer rate and the potential energy as $\Gamma
_{contrast} \equiv ( |\Gamma_{+}|-|\Gamma_{-}| )/( |\Gamma_{+}|+|\Gamma_{-}| )$ and
$\Delta_{contrast} \equiv ( |\Delta_{+}|-|\Delta_{-}| )/( |\Delta_{+}|+|\Delta_{-}| )$,
where $\Gamma_{\pm}$ and $\Delta_{\pm}$ are $\Gamma_{AB}$ and $\Delta_{AB}$
for $\mathbf{\hat{u}}_{A}=\mathbf{\hat{u}}_{A}^{\pm}$. They describe the polarization dependent property for the DDI. If
$\Gamma_{contrast}$ is close to 1 or -1, there is a
huge difference for the transfer rate between $\mathbf{\hat{u}}_{A}=\mathbf{\hat{u}}_{A}^{+}$\ and $\mathbf{\hat{u}%
}_{A}=\mathbf{\hat{u}}_{A}^{-}$.  The same is true for $\Delta_{contrast}$. Figure 2(a) and 2(b) show $\Delta
_{contrast}$\ and $\Gamma_{contrast}$. The insets show that over a broad frequency range, $\Delta_{contrast}$\ and $\Gamma_{contrast}$ are nearly -1, which
mean that atom B with the $x$-polarized transition dipole is much more strongly
coupled to atom A with the left hand circularly polarized dipole $\mathbf{\hat
{u}}_{A}^{-}$\ than with the right hand circularly polarized dipole
$\mathbf{\hat{u}}_{A}^{+}$. This can be seen clearly in Fig. 2 (c) and Fig. 2
(d). For example, $|\Gamma_{+}|/|\Gamma_{-}|=3.34\times10^{-6}$ with $|\Gamma_{-}|=8987 \Gamma_0$\ and
$|\Gamma_{+}|=0.03\Gamma_0$\ for $\omega=2.937eV$.%
\begin{figure}[tbph]
\includegraphics[width=8.5cm]{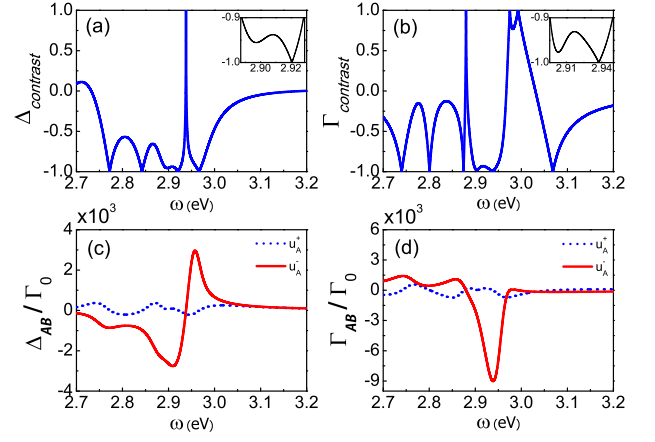}
\caption{(Color online). Polarization-dependent characteristics for the
potential energy $\Delta_{AB}$\ and the transfer rate $\Gamma_{AB}$\ with
$\mathbf{\hat{u}}_{B}=\hat{\mathbf{e}}_{x}$\ and $\mathbf{\hat{u}}_{A}=\mathbf{\hat{u}}%
_{A}^{\pm}=(\hat{\mathbf{e}}_{x}\pm i\hat{\mathbf{e}}_{y})/\sqrt{2}$. (a) The contrast of
$\Delta_{AB}$\ and (b) $\Gamma_{AB}$\ for the two different polarizations of atom A, with $\Delta_{contrast}=(|\Delta_{+}|-|\Delta
_{-}|)/(|\Delta_{+}|+|\Delta_{-}|)$\ and $\Gamma_{contrast}=(|\Gamma
_{+}|-|\Gamma_{-}|)/(|\Gamma_{+}|+|\Gamma_{-}|)$. $\Delta_{\pm}$\ and
$\Gamma_{\pm}$\ are $\Delta_{AB}$\ and $\Gamma_{AB}$\ with $\mathbf{\hat{u}%
}_{A}=\mathbf{\hat{u}}_{A}^{\pm}$. The insets show that over a broad frequency
range, $\Delta_{contrast}$\ and $\Gamma_{contrast}$\ are nearly -1 which means
that the resonance DDI is much stronger for $\mathbf{\hat{u}}_{A}=\mathbf{\hat{u}%
}_{A}^{-}$\ than for $\mathbf{\hat{u}}_{A}=\mathbf{\hat{u}}_{A}^{+}$. This can be clearly seen in (c) and (d) where the
normalized $\Delta_{AB}$ and $\Gamma_{AB}$ are shown for $\mathbf{\hat{u}}_{A}=\mathbf{\hat{u}}_{A}^{\pm}$. The position of atom A is $\mathbf{r}_{A}=(19nm,11nm)$.}
\label{fig2}
\end{figure}

The above results clearly demonstrate that both of the transfer rate and the
potential energy have the great polarization selectivity and deterministic
coupling can take place over a wide frequency range for atoms located at some
certain spatial positions. However, in practice, there will be some degree of
indeterminacy for the positions of the atoms or the radius of the ANS. In order
to show these in greater detail, we make a slight change to both the position of atom A and the radius of the ANS. Figure 3(a) and 3(b) show $\Delta_{contrast}$ and $\Gamma_{contrast}$ for $a=18nm$. Results for four slightly different positions are shown, where the blue (solid) is for $r_{A}=(17nm,12nm)$, the red (dot) is for $r_{A}=(17nm,11nm)$, the black (short dash dot) is for $r_{A}=(16nm,11nm)$, and the dark cyan (short dash) is for $r_{A}=(17nm,13nm)$. We find that deterministic coupling can take place over a wide frequency range (about $0.03eV$) with some variation for the positions of the atoms and the radius of the ANS. Figure 3(c) and 3(d) are for $a=20nm$ and show similar results.
\begin{figure}[tbph]
\includegraphics[width=8.5cm]{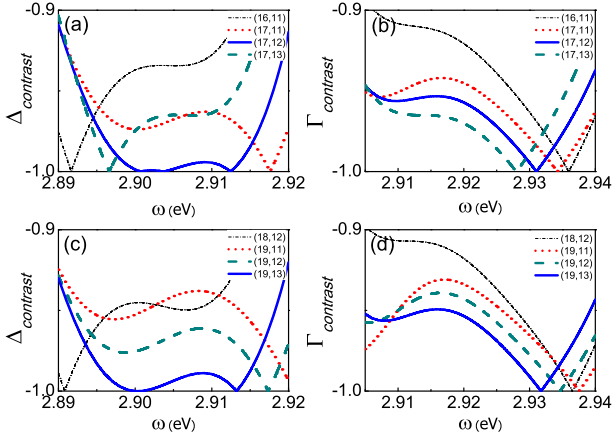}
\caption{(Color online). The stabilization of the nearly one $\Delta_{contrast}%
$\ and $\Gamma_{contrast}$ for different positions of atom A and the radius of
the ANS $a$. (a) and (b) are for $a=18nm$. The blue (solid) is for $r_{A}=(17nm,12nm)$, the red (dot) is for $r_{A}=(17nm,11nm)$, the black (short dash dot) is for $r_{A}=(16nm,11nm)$, and the dark cyan (short dash) is for $r_{A}=(17nm,13nm)$. At a wide frequency range, $\Delta_{contrast}%
$\ and $\Gamma_{contrast}$\ are nearly -1. Similar phenomenon are found in (c)
and (d) for $a=20nm$\ where the black (short dash dot) is for $r_{A}=(18nm,12nm)$, the red (dot) is for $r_{A}=(19nm,11nm)$, the blue (solid) is for $r_{A}=(19nm,13nm)$, and the dark cyan (short dash) is for $r_{A}=(19nm,12nm)$. }
\label{fig3}
\end{figure}

Besides of the above deterministic coupling where only one kind of the two
polarized transition dipole of atom A takes
effect over a wide frequency range, there is another interesting phenomenon
where a sudden change takes place within a narrow frequency range for two
different types of transition dipole ($\mathbf{\hat{u}}_{A}^{\pm}=(\hat{e}_{y}\pm i\hat{e}_{z})/\sqrt{2}$). This
can be clearly seen in Fig. 4(a), where $\Gamma_{contrast}=1$ for
$\omega=2.936eV$ and $\Gamma_{contrast}=-1$\ for $\omega=2.95eV$. Figure 4(b)
shows the normalized transfer rate for $\mathbf{\hat{u}}_{A}=\mathbf{\hat{u}%
}_{A}^{\pm}$. We get $\Gamma_{+}/\Gamma_{0}=6520$ and $\Gamma_{-}/\Gamma
_{0}=0$\ for $\omega=2.936eV$. Accordingly, $\Gamma_{+}/\Gamma_{0}%
=0$\ and $\Gamma_{-}/\Gamma_{0}=-3883$\ for $\omega=2.95eV$. This means
that atom B can couple to atom A with the polarization changing from
$\mathbf{\hat{u}}_{A}=\mathbf{\hat{u}}_{A}^{+}$\ to $\mathbf{\hat{u}}%
_{A}=\mathbf{\hat{u}}_{A}^{-}$\ by fine tuning the transition frequency. Different from the previous results where $|\Gamma_{+}|$ ($|\Delta_{+}|$) are much more weak than $|\Gamma_{-}|$ ($|\Delta_{-}|$), here, both $|\Gamma_{+}|$ and $|\Gamma_{-}|$ can be large depending on the transition frequency.
This flip-flop property can be utilized in the quantum control of the DDI.%

\begin{figure}[tbph]
\includegraphics[width=8.5cm]{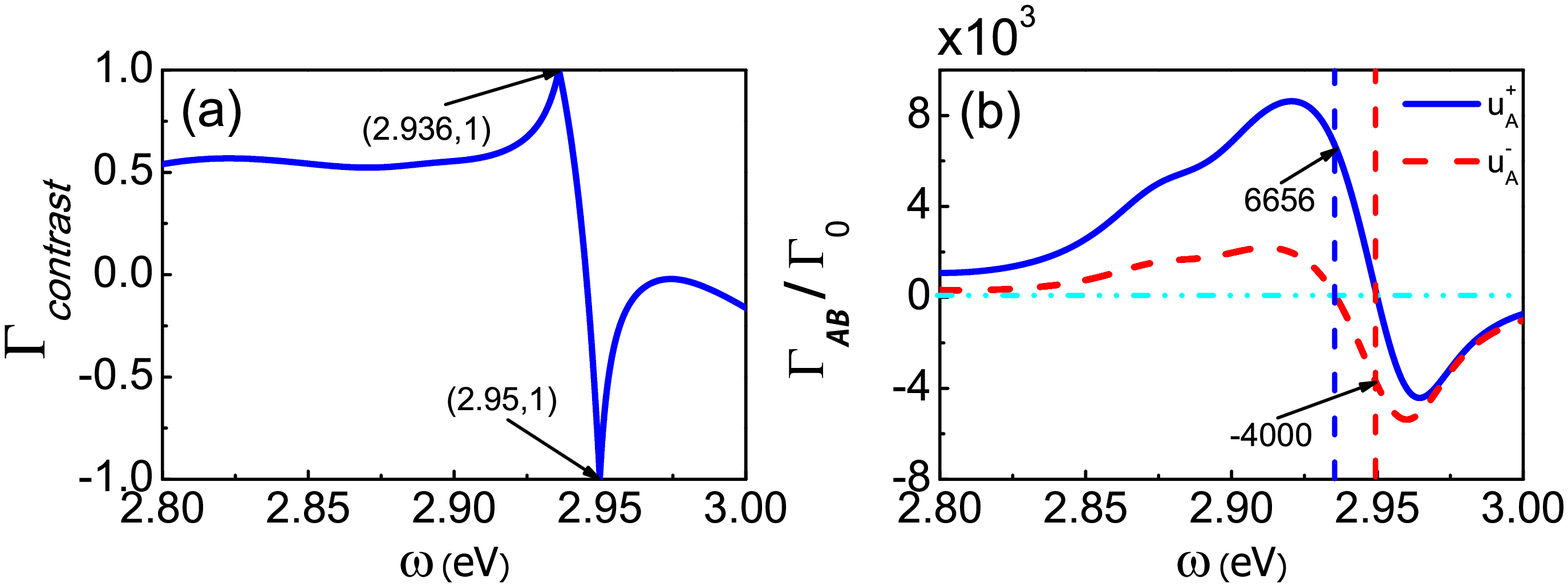}
\caption{(Color online). Switch from $\mathbf{\hat{u}}_{A}=\mathbf{\hat{u}}%
_{A}^{+}$\ to $\mathbf{\hat{u}}_{A}=\mathbf{\hat{u}}_{A}^{-}$\ for atom A to
couple with atom B with $\mathbf{\hat{u}}_{B}=(\hat{e}_{x}+ i\hat{e}_{z})/\sqrt{2}$  by fine tuning the transition frequency. (a) $\Gamma
_{contrast}$ and (b) $\Gamma_{AB}/\Gamma_{0}$\ versus frequency for
$\mathbf{\hat{u}}_{A}=\mathbf{\hat{u}}_{A}^{\pm}=(\hat{e}_{y}\pm i\hat{e}%
_{z})/\sqrt{2}$. The conversion takes place within a narrow frequency range.
The position of the atom A is $r_{A}=(19nm,9nm)$.}
\label{fig4}
\end{figure}

Even though the above results are got for some certain position of atom A, our
numerical tests show that there are other positions to demonstrate the similar
phenomena. Another interesting case is for atom A locating on the $x$-axis as
well. In this situation, owing to the symmetry of our system about the $x$-axis,
we get $(G_{ij}=0)$ $\left(  i,j=x,y,z\right) $ for $i\neq j$ and
$G_{yy}=G_{zz}$. Thus, a linearly polarized transition dipole can only
interact with another dipole which has component along the same direction.
This is simple and we consider circularly polarized transition dipole moment.
Figure 5(a) and 5(b) show that $\Gamma_{contrast}$\ and $\Delta_{contrast}%
$\ are always 1 regardless of the transition frequency, where we set
$\mathbf{\hat{u}}_{A}=\mathbf{\hat{u}}_{A}^{\pm}=(\hat{\mathbf{e}}_{y}\pm i\hat{\mathbf{e}}%
_{z})/\sqrt{2}$ and $\mathbf{\hat{u}}_{B}=(\hat{\mathbf{e}}_{y}+ i\hat{\mathbf{e}}_{z}%
)/\sqrt{2}$ with $\mathbf{r}_{A}=(-25nm,0)$ and $\mathbf{r}_{B}=(22nm,0)$. The origin of
this character stems from the fact that $G_{yy}=G_{zz}$. If we substitute
$\mathbf{\hat{u}}_{A}$\ and $\mathbf{\hat{u}}_{B}$ into Eq. (\ref{eq:gamadeltazhongyao}) and make use of the result that $G_{yy}=G_{zz}$, we get $\Gamma_{-}=0$\ and
$\Delta_{-}=0$\ . This effect has been numerical verified in Fig. 5(c) and
5(d). By the same procedure, we can easily prove that atom A can couple to atom B
only when $\mathbf{\hat{u}}_{A}^{\ast}\cdot\mathbf{\hat{u}}_{B}\neq 0$ for both
transition dipole moments polarized in the $yz$-plane.
\begin{figure}[tbph]
\includegraphics[width=8.5cm]{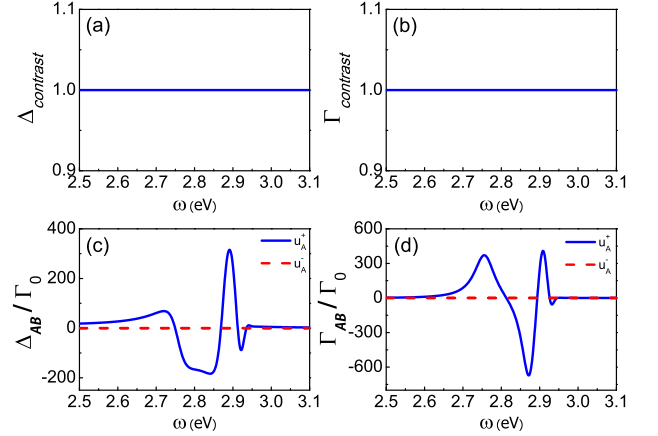}
\caption{(Color online). The transfer rate and potential energy for both atoms
located on the $x$-axis with $\mathbf{\hat{u}}_{A}=\mathbf{\hat{u}}_{A}^{\pm
}=(\hat{e}_{y}\pm i\hat{e}_{z})/\sqrt{2}$\ and $\mathbf{\hat{u}}_{B}=(\hat
{e}_{y}+ i\hat{e}_{z})/\sqrt{2}$. (a) and (b) are for $\Delta_{contrast}$
and $\Gamma_{contrast}$\ . (c) and (d) are for $\Delta_{AB}/\Gamma_{0}$ and
$\Gamma_{AB}/\Gamma_{0}$. Here $\mathbf{r}_{A}=(-25nm,0)$ and $\mathbf{r}_{B}=(22nm,0)$.}
\label{fig5}
\end{figure}

\emph{Conclusion.} In this work, we have shown that the polarization takes great effect on the DDI. Theoretically, we have proposed a formalism to treat the DDI between two polarized dipoles, where the transfer rate and the potential
energy are expressed by the Green function (Eq. (\ref{eq:gamadeltazhongyao})). Different from the real dipole case where the coherent part $\Gamma_{ij}$ (incoherent part $\Delta_{ij}$) is related to the imaginary (real) part of the Green function, $\Gamma_{ij}$ ($\Delta_{ij}$) is related to both the real part and the imaginary part of the Green function for complex transition dipole moment. By utilizing this general expression and the analytic
Green function of the ANS, we have found that an atom with linear transition dipole couples much stronger to the other atom with left circularly polarized transition dipole than with right circularly polarized transition dipole. We have shown that the above polarization selective DDI can take place over a wide frequency range (about $0.03eV$) and it is robust against variation of the atom's position and the radius of the ANS. Besides the stable deterministic coupling, we have also found that a sudden change takes place within a narrow frequency range (about $0.014eV$), where the contrast of the transfer rate for $\mathbf{\hat{u}}_{A}=(\hat{e}_{y}\pm i\hat{e}_{z})/\sqrt{2}$ changes from 1 to -1. We have also investigated the case where both of the two dipoles are locating on the \emph{x}-axis and found that an atom with right circularly polarized transition dipole couples to the atom with the same polarization and not to the left circularly polarized transition dipole. We have proved that this deterministic polarization selective characteristic for DDI is symmetry protected.

\begin{acknowledgments}
 This work was financially supported by the National Natural Science Foundation
of China (Grants No. 11464014, 11347215, 11564013, 11402096, 11464013).
\end{acknowledgments}

\bibliography{introductionbib}

\end{document}